\documentclass{elsart}

\usepackage{natbib}
\usepackage{epsfig}
\usepackage{amssymb}
\usepackage{graphicx}
\usepackage{rotating}

\begin{document}

\begin{frontmatter}

  \title{A Model Grid for the Spectral Analysis of X-ray Emission in Young Type Ia Supernova Remnants}
  
  \author{Carles Badenes\corauthref{cor1}}
  
  \address{Department of Physics and Astronomy, Rutgers University, 136 Frelinghuysen Rd., 
    Piscataway NJ 08854-8019}
  
  \corauth[cor1]{Corresponding author. E-mail: badenes@physics.rutgers.edu}
  
  \author{Eduardo Bravo\thanksref{label1}}
  
  \address{Departament de F\'{i}sica i Enginyeria Nuclear, Universitat Polit\`{e}cnica de Catalunya, Diagonal 647, 
    Barcelona 08028, Spain}
  
  \author{Kazimierz J. Borkowski}
  \address{Department of Physics, North Carolina State University, Box 8202, Raleigh NC 27965-8202} 
  
  \thanks[label1]{Also at Institut d'Estudis Espacials de Catalunya, Campus UAB, Facultat de Ci\`{e}ncies. Torre C5. Bellaterra, 
  Barcelona 08193, Spain}
 
  \begin{abstract}
    We address a new set of models for the spectral analysis of the X-ray emission from young, ejecta-dominated Type Ia
    supernova remnants. These models are based on hydrodynamic simulations of the interaction between 
    Type Ia supernova explosion models and the surrounding ambient medium, coupled to self-consistent ionization and electron 
    heating calculations in the shocked supernova ejecta, and the generation of synthetic spectra with an appropriate spectral 
    code. The details are provided elsewhere, but in this paper we concentrate on a specific class of Type Ia explosion
    models (delayed detonations), commenting on the differences that arise between their synthetic X-ray spectra under a variety
    of conditions. 
  \end{abstract}

  
\end{frontmatter}

\section{A library of synthetic X-ray spectra for Type Ia supernova remnants} \label{sec:Intro}

There is little doubt that the X-ray spectra of the supernova remnants (SNRs) originated by Type Ia supernovae (SNe) 
contain important information regarding the physical mechanism behind these intriguing explosions. Many of these spectra
are dominated by emission lines from the shocked supernova ejecta, where the products of thermonuclear nucleosynthesis 
can be clearly seen. This information, however, is not easy to extract, mainly due to the fact that the X-ray spectra
of ejecta-dominated SNRs are very difficult to interpret and analyze. 

We have developed a library of synthetic spectra for the analysis of the X-ray emission from young, ejecta-dominated Type Ia SNRs. This 
library is generated using one dimensional hydrodynamic simulations of the interaction of Type Ia SN explosion models with a uniform
ambient medium (AM) and detailed calculations of the ensuing nonequilibrium ionization and electron heating processes in 
the shocked ejecta. Once the state of the shocked plasma is characterized in this way, synthetic X-ray spectra can be generated
with an appropriate spectral code. The relevant techniques are outlined in \citet{badenes03:xray} and \citet{badenes04:xray}
(henceforth, Paper I and Paper II). A detailed comparison between these synthetic spectra and the observations
of the Tycho SNR will be the subject of a forthcoming paper (Badenes et al., in preparation). In the 
present work, we concentrate on a particular class of Type Ia models, the delayed detonations. First, we review the 
state of the art of the delayed detonation paradigm and discuss the differences between the models in this class. Then, we present 
a series of synthetic X-ray SNR spectra calculated from different delayed detonation models and examine the properties of the 
X-ray emission obtained in each case. Our aim is to illustrate the potential of our 
synthetic spectra by focusing on a class of models which are essentially similar to one another, but whose X-ray spectra in the
SNR phase show important differences that are well within the resolving power of the current X-ray observatories.

\section{Delayed detonation: a phenomenological model for thermonuclear supernova explosions} \label{sec:DDT}

The delayed detonation (DDT) paradigm was proposed in \citet{khokhlov91:ddt} as an alternative to pure deflagrations,
which had been the preferred model for thermonuclear supernovae until then. In this paradigm, the burning front inside 
the white dwarf (WD) starts to propagate as a subsonic flame, but at a given point the flame makes a transition to the 
supersonic regime, and a detonation ensues that burns the rest of the WD. This kind of Type Ia SN models is  
considered the most successful paradigm for reproducing the light curves and spectra of Type Ia SNe \citep{hoeflich96:nosubCh},
but it has to be kept in mind that it does not provide a self-consistent explanation for the physics of Type Ia SNe. 
An important issue that still needs to be clarified is the physical mechanism for the transition from deflagration
to detonation, which is always induced artificially in all DDT models. Another issue is the fact that the
vast majority of the published DDT models are calculated in 1D, and are becoming obsolete compared to modern
3D simulations. Recently, the interest in the DDT paradigm has been rekindled as a means to avoid the thorough mixing 
of fuel and ashes that seems to be unavoidable in 3D deflagrations 
\citep{reinecke02:Ia3D,gamezo03:Ia3D,travaglio04:3D,garcia-senz03:Ia3D}. 
The first DDT models in 3D have begun to appear in the literature \citep{garcia-senz03:DDT3D,gamezo05:DDT3D}, 
but the results of these calculations do not converge, and we are still far from a unified picture of the DDT paradigm in 
3D. Until the ability of 3D calculations to reproduce light curves and spectra of Type Ia SNe is fully established
\citep[see][]{baron03:detectability,bravo04:3D_review}, 1D DDT models remain the most successful Type Ia SN models,
despite all their misgivings.

\begin{table}
  \centering
  \begin{tabular}{ccccccccccc}
    \hline 
    Model & $\iota$ & $\rho_{tr}$ &
    $E_{k}$ & $M_{Fe}$ &
    $M_{C+O}$ & $M_{Si}$ & $M_{S}$ & $M_{Ar}$ &
    $M_{Ca}$ \\
    & & [$\mathrm{g\cdot cm^{-3}}$] &
    [$10^{51}\mathrm{erg}$] & [$\mathrm{M_{\odot}}$] &
    [$\mathrm{M_{\odot}}$] & [$\mathrm{M_{\odot}}$] & [$\mathrm{M_{\odot}}$] & [$\mathrm{M_{\odot}}$] &
    [$\mathrm{M_{\odot}}$] \\
    \hline
    \hline 
    DDTa & 0.03 & $3.9\cdot10^{7}$ &
    1.40 & 1.03 & 0.04 & 0.087 & 0.071 & 0.019 & 0.022 \\
    DDTbb & 0.01 & $2.5\cdot10^{7}$ &
    1.31 & 0.99 & 0.05 & 0.10 & 0.084 & 0.022 & 0.027 \\
    DDTc & 0.03 & $2.2\cdot10^{7}$ &
    1.16 & 0.80 & 0.12 & 0.17 & 0.13 & 0.033 & 0.038 \\
    DDTe & 0.03 & $1.3\cdot10^{7}$ & 
    0.94 & 0.56 & 0.19 & 0.25 & 0.19 & 0.046 & 0.054 \\
    \hline 
  \end{tabular}

  \caption{Characteristics of the four DDT models, reproduced from Paper I and paper II. The
    total ejected mass is $M_{ej}=1.37\,\mathrm{M_{\odot}}$ in all cases. $E_{k}$ is the kinetic energy. \label{tab-1}}

\end{table}

\begin{figure}

  \centering
  \includegraphics[scale=0.7]{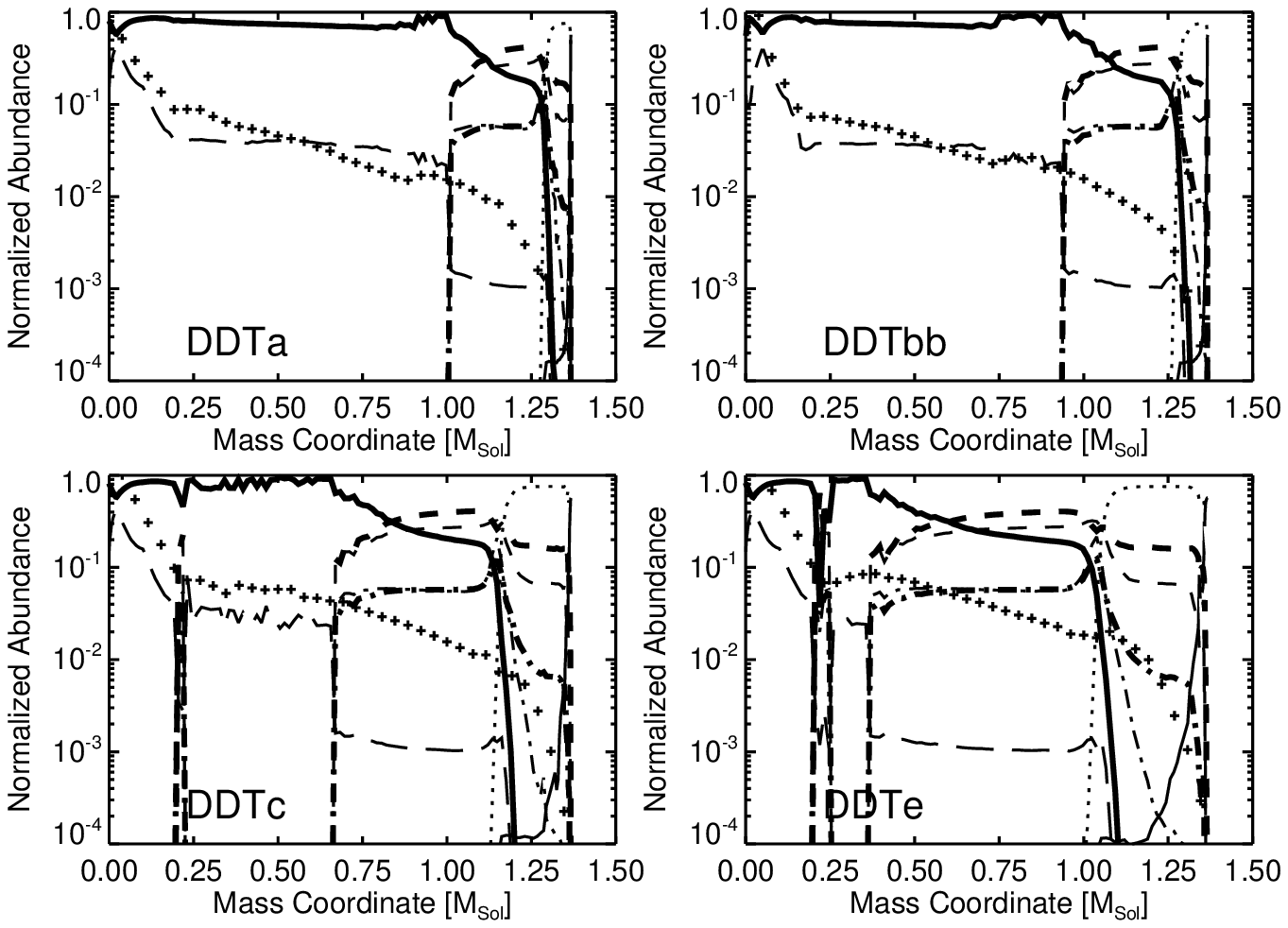} 
  \includegraphics[scale=0.7]{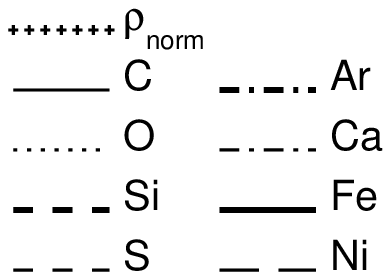}

  \caption{Composition and density profiles for the DDT Type Ia SN explosion models from our grid (reproduced from Paper I
    and Paper II). The abundances represented are number abundances after the decay of all short lifetime 
    isotopes. The density profiles are represented at $t=10^{6}$s after the SN explosion, normalized by 
    $\rho_{n}=10^{-11}\mathrm{g \cdot cm^{-3}}$. \label{fig-1}}

\end{figure}

In Table \ref{tab-1} and Figure \ref{fig-1}, we provide the characteristics of four DDT models in our grid of Type Ia SN
explosions. These models can be found in Papers I and II, they are just reproduced here for the convenience of the reader.
The fundamental properties of these models, like the chemical composition profile, the density profile and 
the total nucleosynthetic yields, depend on the main parameter involved in the calculations, $\rho_{tr}$, which represents
the density inside the WD at which the burning front is forced to make the transition from flame to detonation. Models with higher 
values of $\rho_{tr}$, like DDTa, produce more energetic explosions, with a higher amount of $^{56}$Ni (which later decays 
to $^{56}$Fe), while models with lower values of $\rho_{tr}$, like DDTe, produce less $^{56}$Ni but more intermediate mass 
elements (IMEs), like Si, S, Ar and Ca. This is explained because for a lower value of $\rho_{tr}$, the detonation propagates 
into a WD that has expanded more during the deflagration phase, and therefore the burning front departs the region of nuclear 
statistic equilibrium sooner, and the IMEs take over the chemical composition profile at a smaller radius. The speed of the flame
during the deflagration phase, which is controlled by the parameter $\iota$, has a much smaller impact on the final
outcome. For more details on the parameters of DDT models and the way the calculations were carried out, see Paper I 
and Paper II.

\section{The signature of $\rho_{tr}$ in the X-ray spectra of SNRs} \label{sec:DDT-SNR}

Within our simulation scheme, the X-ray spectrum from the ejecta in the SNR is determined by four factors: the SN explosion
model, the density of the AM ($\rho_{AM}$), the age of the SNR ($t$), and the amount of internal energy that is deposited 
in the electrons at the reverse shock ($\beta$, defined as the ratio of electron to ion postshock specific internal energy, 
see Paper II). In Paper I and Paper II, the impact that each of these four factors has on the X-ray spectra 
of Type Ia SNRs is analyzed in a general context. Here, we will focus on DDT models and on the imprint of 
$\rho_{tr}$.
 
\begin{figure}

  \centering
  \includegraphics[scale=0.66,angle=90]{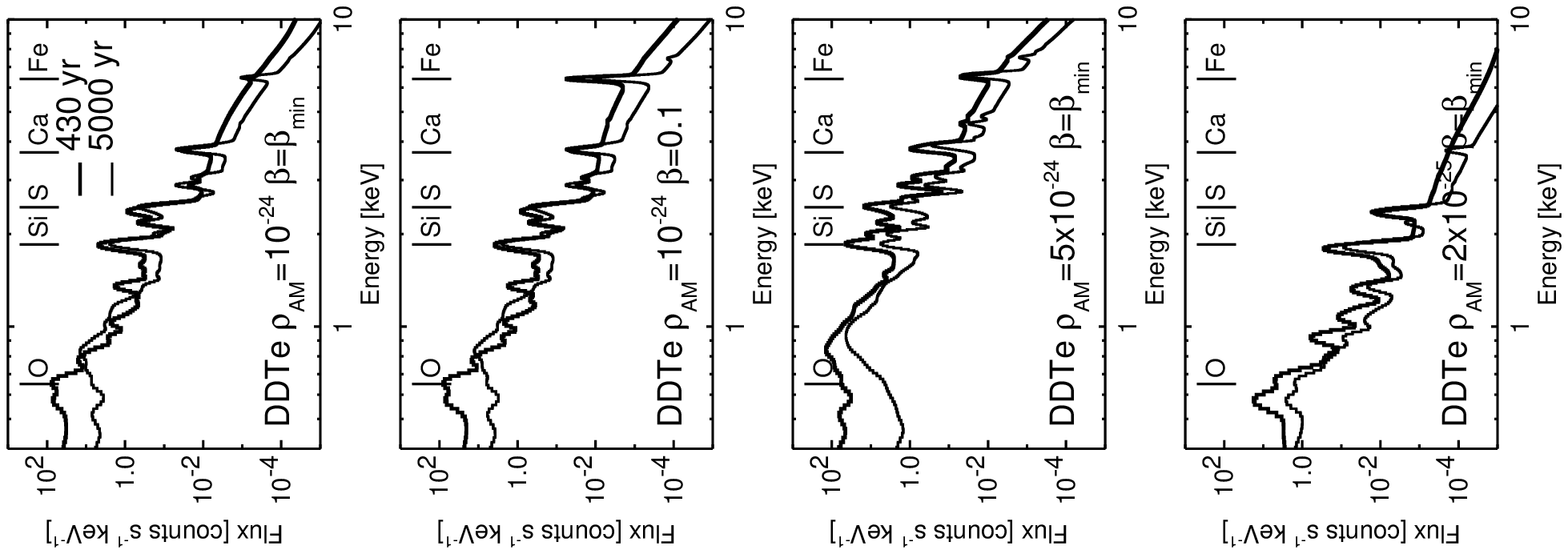}
  \includegraphics[scale=0.66,angle=90]{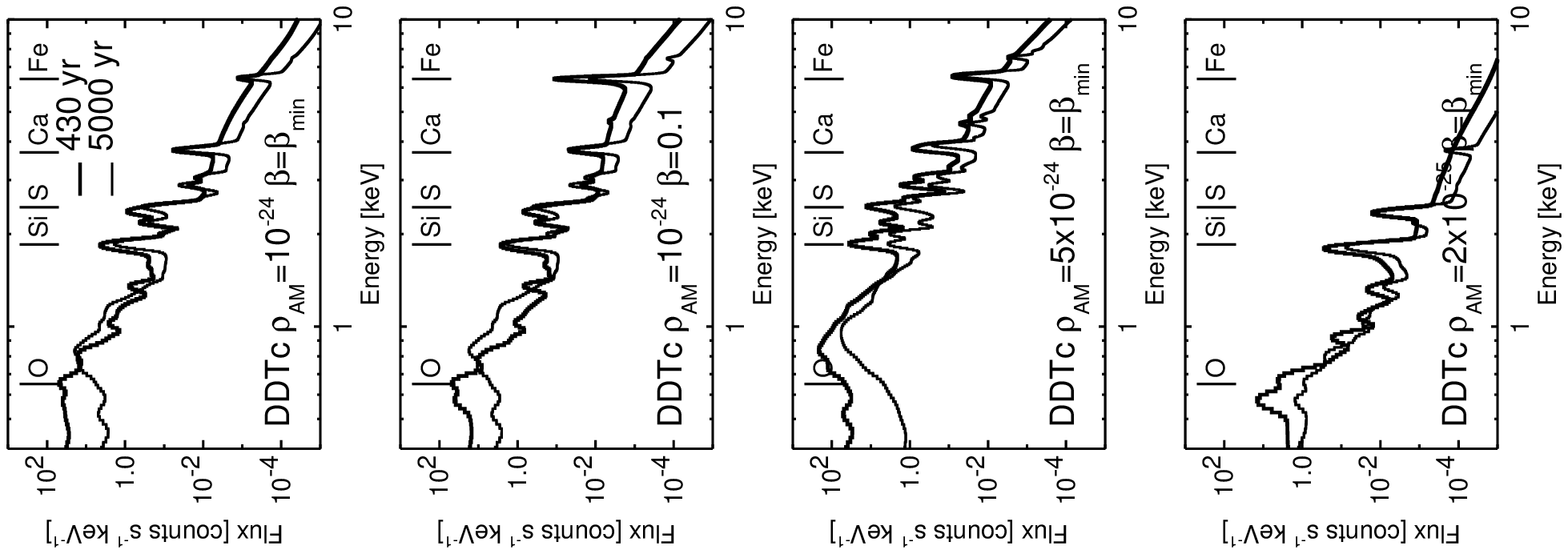} 
  \includegraphics[scale=0.66,angle=90]{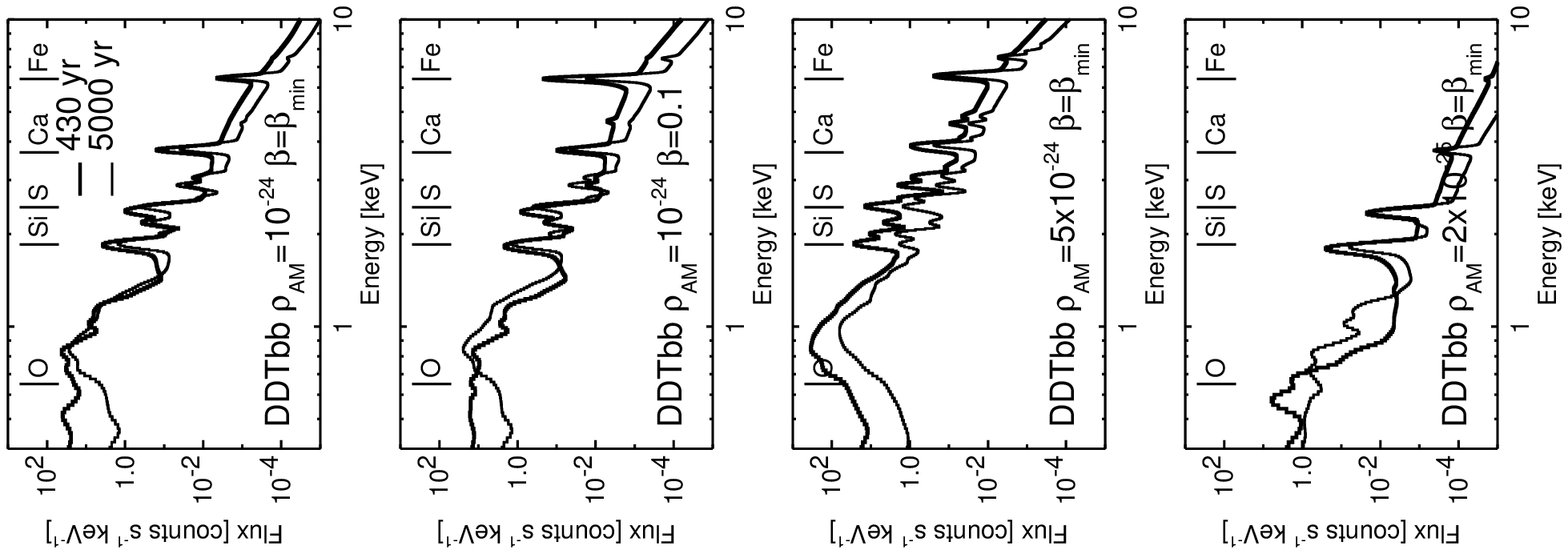} 
  \includegraphics[scale=0.66,angle=90]{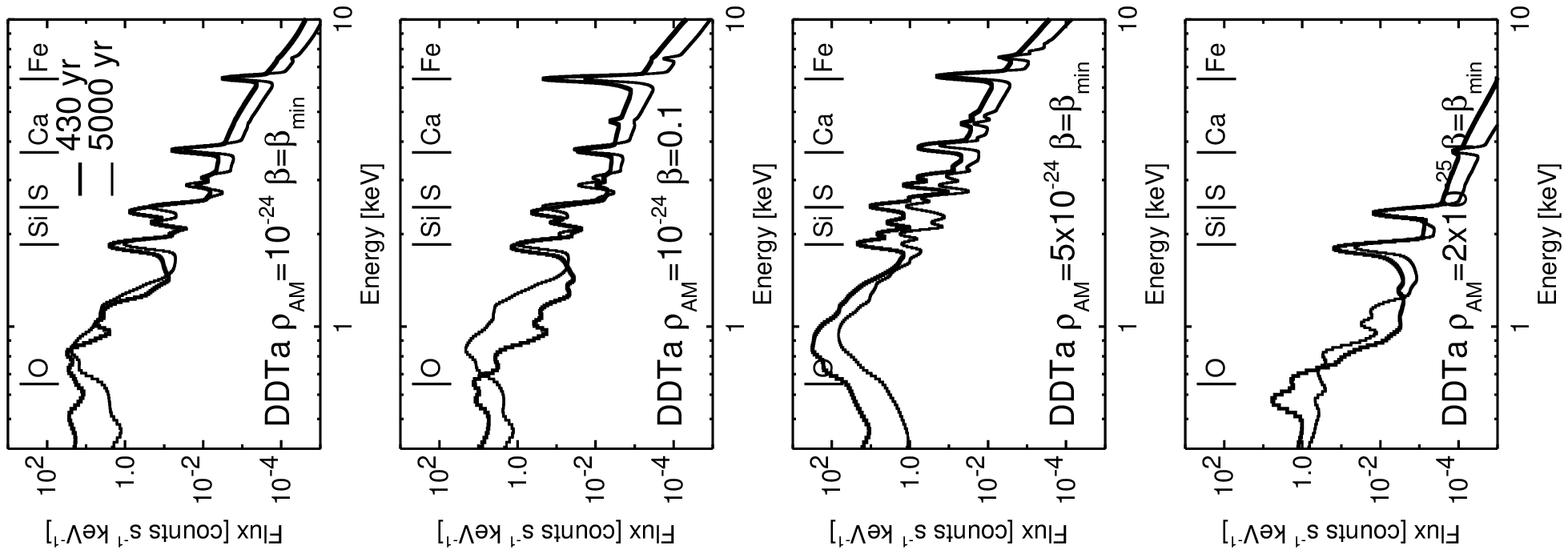} 

  \caption{Unabsorbed X-ray spectra from the shocked ejecta of the four DDT explosion models DDTa, DDTbb, DDTc, and DDTe, given 
    at $t$=430 yr (thick solid lines) and 5,000 yr (thin dotted lines). The values of $\rho_{AM}$ are: $10^{-24}\, g \cdot cm^{-3}$ 
    (panels in the first and second rows), $\rho_{AM}=5 \cdot 10^{-24}\, g \cdot cm^{-3}$ (panels in the third row), 
    and $\rho_{AM}=2 \cdot 10^{-25}\, g \cdot cm^{-3}$ (panels in the fourth row). The value of $\beta$ is 0.1 for the panels in
    the second row and $\beta_{min}$ elsewhere. The spectra have been convolved with the response of the {\it XMM-Newton} EPIC
    MOS1 camera. The Ly$\alpha$ line of O, and the K$\alpha$ lines of Si, S, Ca and Fe have been marked for clarity. Note that 
    the spectral code has no atomic data for Ar.\label{fig-2}}

\end{figure}

In the top two rows of Figure \ref{fig-2}, we plot the synthetic ejecta spectra for the four DDT models of Table \ref{tab-1} at 
SNR ages of 430 yr (the age of Tycho's SNR) and 5000 yr, for $\rho_{AM}=10^{-24}\,\mathrm{g \cdot cm^{-3}}$. In the first row of
panels, no collisionless electron heating has been introduced at the reverse shock ($\beta=\beta_{min}$, equal to 
the ratio of electron to ion masses, see Paper II for a discussion). 
In this case, the 
signature of $\rho_{tr}$ is easy to identify, as emission in the Fe L and Fe K$\alpha$ complexes varies from very prominent 
(DDTa, high $\rho_{tr}$) to hardly detectable (DDTe, low $\rho_{tr}$). Despite the fact that the masses of Si and S in the 
ejecta vary by a factor 2.5 between the models, the emission in the Si and S He$\alpha$ and He$\beta$ lines does not change 
much from DDTa to DDTe. Note that the Fe emission increases at late times for DDTe and DDTc, but is always much lower than in 
DDTa or DDTbb. A remarkable feature of the models with low $\rho_{tr}$ is the enhanced emission from O, Ne and Mg at early times 
(Ne and Mg are synthesized in minor amounts in the O-rich regions of DDTc and DDTe, and they are only revealed if the Fe L
emission is low). If collisionless electron heating at the reverse shock is introduced ($\beta=0.1$, 
second row of panels in Figure \ref{fig-2}), the flux in the Fe K$\alpha$ complex, which probes material 
at higher temperatures than Fe L, is enhanced, 
and the thermal continuum is affected in all cases (this is hard to see in the plots, but becomes apparent if the bremsstrahlung 
temperature is fitted). In model DDTa, the Fe L emission is significantly reduced, but only at early times, while the excess electron 
temperature affects most of the shocked ejecta (see Paper II). Given the stratified structure of DDT models, 
only Fe is severely affected by collisionless electron heating, at least for the values of $t$ that we explore here. The signature 
of $\rho_{tr}$ in the Fe emission is harder to identify for $\beta=0.1$, but the presence of O, Ne and Mg is still noticeable 
for the models with low $\rho_{tr}$.

In the bottom two rows of Figure \ref{fig-2}, the spectra of the four DDT models are plotted for different 
values of $\rho_{AM}$. At higher AM
densities ($\rho_{AM}=5 \cdot 10^{-24}\,\mathrm{g \cdot cm^{-3}}$), the SNR models are more evolved at a given age, 
the emitted X-ray flux is higher, and the ionization state of all 
the elements in the spectrum is much more advanced. This tends to make the differences between models smaller: Fe L emission, 
for instance, is now found in all cases, and this makes the O, Ne and Mg emission in DDTc and DDTe harder to detect.
The only surviving signature of $\rho_{tr}$ is the higher flux in the Fe K$\alpha$ blend for DDTa and DDTbb, but now the 
difference is much less noticeable. At lower AM densities ($\rho_{AM}=2 \cdot 10^{-25}\,\mathrm{g \cdot cm^{-3}}$), on 
the other hand, the SNRs are much less evolved and the emitted flux is much lower. Fe emission is virtually absent in 
all cases, and the imprint of $\rho_{tr}$ rests again with the O, Ne and Mg emission.

In previous works (Paper I and Paper II), we showed that the X-ray spectra of SNR models obtained from different 
kinds of Type Ia explosions are very different, and that it is possible to use the SNR observations of modern
satellites like {\it Chandra} and {\it XMM Newton} to probe the physics of Type Ia SNe. In this short paper, we have focused on 
delayed detonation models, and we have showed that, although these models are very similar to one another, the X-ray 
spectra that we obtain still show important differences. In a forthcoming paper (Badenes et al., in preparation), we will
make a detailed comparison between our spectra and the observations of the Tycho SNR, and show to what degree
can these differences be identified in a real case.

\end{document}